\documentclass[12pt]{article}

\usepackage{latexsym}
\usepackage{amssymb}
\usepackage{amsmath}

\def\qti {\widetilde{q}}
\def\tti {\widetilde{t}}
\def\mh {\widehat{m}}
\def\bbbone {{\mathchoice {\rm 1\mskip-4mu l} {\rm 1\mskip-4mu l}
{\rm 1\mskip-4.5mu l} {\rm 1\mskip-5mu l}}}

\def\Mo {\widehat{M}}

\begin{document}

\noindent                               
\begin{titlepage}                   

\title{\Large \bf Branes in the  plane wave background with  gauge field condensates}

\author{{\Large Tako Mattik}\thanks{mattik@itp.phys.ethz.ch}\\[4ex]   Institut f\"ur Theoretische Physik
\\ ETH Z\"urich\\CH-8093 Z\"urich\\ Switzerland}
\date{24.1.2005}
\maketitle \abstract{Supersymmetric branes in the plane wave
background  with additional constant magnetic fields are studied
from the world-sheet point of view. It is found that in contradistinction to
flat space,  boundary condensates on some maximally supersymmetric 
 branes necessarily  break  at least some  supersymmetries.  The  maximally supersymmetric cases with condensates  are shown  to be in one to one correspondence with the previously classified class II  branes.}

\thispagestyle{empty}
\end{titlepage}

\section{Introduction}\label{intro}   Since  the discovery of the  plane wave  
 in \cite{blau}   as another maximally supersymmetric
background of type II B superstring theory and  the explicit
quantization of strings on it  in \cite{metsaev1,metsaev2}, this solution
has been intensively used
in the study of the   gauge-gravity (AdS/CFT) correspondence via the BMN proposal \cite{berenstein}. For reviews see for example \cite{sadri1,maldacena2, plefka,pankiewicz, russo}. Branes in this 
Ramond-Ramond background have been studied in a
number of papers from different points of view. The probe brane
approach was carried out in  \cite{skentay1,skentay4,bain2},  boundary
states were used  in \cite{billo,bgg,gabgre,oblique} and  open
string theory methods
 were  applied in  \cite{dabholkar, skentay2, skentay3, bain, chu}. Closely related setups were considered for example  in \cite{alday, sadri2, alishahiha, hyun, kim}.\\[1ex]
In this paper we study branes in the plane wave  with nonzero
gauge field  condensates ${\cal F}^{IJ}, {\cal F}^{I+}$ from the
world-sheet point of view. We derive the conserved (dynamical)
supersymmetries, calculate open string partition functions and
prove the equivalence with the
 closed string cylinder diagrams, expressed by boundary state
 overlaps.\\
It will be shown that in contradistinction to the situation in flat Minkowski space  it is  impossible to turn on
  magnetic fields on some   supersymmetric branes  in the plane wave without further reducing
  the  amount of conserved supersymmetries.\\[2ex]  Maximally supersymmetric branes in the plane wave  without boundary condensates   were classified in \cite{skentay2, gabgre} by using  the fermionic gluing matrix $M$  which  is  given as in flat space  by the product of $\gamma$-matrices along the Neumann directions \begin{equation} M=\prod_{I\in{\cal N}} \gamma^I.\end{equation}   The class I ($D_-$) branes are characterised  by $ M\Pi M\Pi=-1$ with $\Pi=\gamma^1\gamma^2\gamma^3\gamma^4$, and are of the form $(r,r+2)$, $(r+2,r)$ with  $r=0,1,2$. Here the notation $(r,s)$ labels the orientation of the branes  with respect to the $SO(4)\times SO(4)$ - background symmetry.\\  The so called class II ($D_+$) branes are characterised  by $ M\Pi M\Pi=+1,$ and consist  of the (0,0) instanton  and the (4,0), (0,4) branes \cite{skentay2, gabgre}.  The last example is special, as these branes  couple  to the nonzero background $F_5$-form flux, leading necessarily  to a nonzero ${\cal F}^{+I}$-worldvolume flux.\\[2ex] 
The gauge field condensates considered in this paper will be seen to  give  rise to new continuous families of maximally supersymmetric D-branes in the plane wave.
  These families  interpolate smoothly  between the Euclidean (class I)   (2,0), (0,2)
 branes
  and the (0,0) instanton and between  the (4,2), (2,4) and the class II (4,0), (0,4) branes. Because of this, the new branes are in one to one correspondence with the previously mentioned class II branes.\\
 On the remaining static
   Euclidean maximally
  supersymmetric  (non-oblique) class I branes, the (3,1), (1,3) branes, constant  magnetic fields reduce the supersymmetry further.  This
  behavior will be seen to be related to the observation from \cite{billo, dabholkar} that the
  (static) class I branes preserve the maximal amount of (dynamical) supersymmetries only when
  placed at the origin of transverse space.\\[1ex]
As a consistency check we consider the equivalence of the open and
 closed string description of the branes under consideration. For
this we extend and connect the results of \cite{bgg, gabgre} by defining
gauge field dependent generalizations
  of the special functions $f^{(m)}_i$ of \cite{bgg} and $g_i^{(m)}$ of \cite{gabgre}. These special functions itself
  were $m$  - dependent
  deformations of the well known  $f$-functions  of Polchinski and Cai from   \cite{polchinski}, which appear in the
   usual   flat space treatment. Similar  deformations of the standard $\Gamma$-function  have appeared  in the context of plane-wave physics in \cite{lucietti1, lucietti2}. \\
The new function $g^{(m)}_2(q, \theta)$, for example, which appears
in the cylinder diagram of (2,0) branes,  interpolates between the $f_2{(m)}$
function from \cite{bgg}  and the $g_2^{(m)}(q)$ function from the
instanton description of \cite{gabgre}.
The angle
variable $\theta\in [0, \pi]$ parameterizes here  the world-volume gauge
field strength ${\cal F}$.\\[3ex]
The situation of gauge condensates extended
along a light-cone direction, that is, only ${\cal F}^{+I}\neq 0$, was addressed previously  in
\cite{skentay1, gabgre, cha} and for the  more general massive backgrounds of \cite{maldacena}  in
\cite{hikida}. Aspects of non-commutativity in the plane wave
  resulting from a nonzero $B$-field background were studied in \cite{chu, kamani}. For other  studies
     of D3 branes with condensates in this background  by supercoset or DBI methods,
 see \cite{metsaev3} and \cite{pal}. Additional
  comments on the related situation  of branes  in the Nappi-Witten background appeared recently in
      \cite{appoll}.\\[1ex] The paper is organized as follows.
      After briefly summarizing general aspects of boundary
      condensates on D-branes and a short review of the situation
      in flat space in section \ref{review}, we will give a closed string boundary state
      description of maximally supersymmetric branes  in the plane wave by considering
      their (dynamical) supersymmetries in section \ref{gluing}. By doing so, we will derive a general 
condition for the gluing matrices of maximally supersymmetric branes, whose solutions will be discussed in section \ref{solutions}.  In the following section \ref{boundarystates},
 certain boundary states and their overlaps will be derived which will finally be compared with the results of a detailed open string treatment from section \ref{open}.
 Certain technical      calculations are deferred to the appendix.

\section{Boundary  Condensates on Dp branes} \label{review}
Let us begin by reviewing some well-known facts about boundary
condensates on D-branes. As described for example in \cite{abouel,
gregut, gregut2, seiberg} and references therein, the introduction
of a boundary condensate with (constant) gauge potential $A$
 on the world volume of a D-brane  leads to the following
 boundary action \begin{equation} \int ds \left( A_I \partial_s X^I-\frac{1}{2}F_{IJ} S\gamma^{IJ}S   \right)
 \end{equation} with the (abelian) field strength  $F=dA$. In the  case of a constant  $B$ - field,  the   bosonic bulk term proportional to  \begin{equation}
  \epsilon^{\alpha\beta}B_{rs}\partial_\alpha X^r \partial_\beta
  X^s\end{equation}
and the corresponding fermionic couplings    become total
derivative
  terms,  and we obtain  for a constant gauge field for which we can set  $A_I =- \frac{1}{2} F_{IJ} X^J$ the
 combined  surface action \begin{equation} \int ds {\cal F}_{IJ} \left(X^{[I}\partial_s X^{J]}-S\gamma^{IJ}S\right).
 \end{equation} Here  we have used   the usual  gauge invariant quantity ${\cal F}=F-B$.\\
  For the Neumann directions this boundary action leads    to the  modified boundary conditions
  \begin{equation}\label{bosbound} \partial_\sigma X^I+{\cal F}_\nu^{IJ} \partial_\tau X^{J}=0\end{equation}
  at $\sigma=0, \pi$. By $\nu=1,2$   possibly different condensates on the branes at $\sigma=0$ and $\sigma=\pi$ are distinguished.
    For simplicity we will, however,  concentrate in the following
  on the case ${\cal F}_1={\cal F}_2$.  As a relation between $\partial_+ X$ and $\partial_- X$  equation
   (\ref{bosbound}) reads \begin{equation}\label{bound1} \left(\partial_+ X^I+N^{IJ} \partial_- X^J\right)=0\;\;\;\;  \end{equation} with
    \begin{equation}\label{boundN2}  N^{IJ}=- \left[\frac{1-{\cal F}}{1+{\cal F}}\right]^{IJ},\;\;\;\; {\cal F}^{IJ}=
    \left[\frac{1+N}{1-N}\right]^{IJ},\end{equation} which is valid for the Neumann directions.
     For the  Dirichlet  directions the condition (\ref{bound1}) will be imposed with $N=1$, that is
      \begin{equation}\left(\partial_+ X^i+ \partial_- X^i\right)=\partial_\tau X^i=0 \label{bosboundd}\end{equation}
       on the boundary\footnote{In this paper we will use upper case letters $I, J...$ for Neumann - and lower case letters
       $i, j...$ for Dirichlet directions.}. \\[2ex]
In \cite{gregut} it  is shown  how to implement a nonzero gauge
field condensate in the light-cone
         gauge  boundary state description of branes in type II string theory in a flat background by using the
conservation of space-time supersymmetries as a guiding principle.
Using boundary states this  conservation is expressed by
\begin{equation}\label{susyflat} \left(Q_{\dot a}+i\eta M_{\dot
a\dot b}\tilde Q_{\dot b}\right) ||B\rangle\rangle=0,\;\;
  \left(Q_{a}+i\eta M_{a b}\tilde Q_{b}\right)
  ||B\rangle\rangle=0\end{equation} for the  16 dynamical and
  16 kinematical supersymmetries preserved by the  (flat)
background.\\ As shown in \cite{gregut}, (\ref{susyflat}) is
  fulfilled when  using the fermionic gluing conditions
  \begin{equation} \left(S^a_n+i\eta M_{ab}\widetilde{S}^b_{-n} \right)   ||B\rangle\rangle=0  \end{equation}
   in addition to the bosonic
  relations (\ref{bosboundd}) and (\ref{bound1}) described above,
  iff  the   orthogonal matrices $N^{IJ}, M_{ab}, M_{\dot a\dot
b}$ appearing in these conditions are related by
  \begin{equation} M_{ab} \gamma^I_{b\dot{c}}
M^t_{\dot{c}\dot{d}}=\gamma^J_{a\dot{d}} N^{JI}.\label{triality1}
\end{equation} This equation expresses, as will be  further explained later on,
a $SO(8)$-triality relation.\\   As there are no further conditions
on the gluing matrices, this means that an arbitrary constant
boundary condensate can be turned on on any even dimensional
world-volume subspace of a supersymmetric brane in flat space without changing
the amount of conserved supersymmetries. \\  For further details
and in particular the derivation of cylinder diagrams as boundary
state overlaps,  the reader is referred to \cite{gregut, gregut2},
for example.

\section{Gluing conditions for maximally supersymmetric branes in the plane wave} \label{gluing}
Now we turn to branes in the plane wave  background, being mainly interested in static maximally supersymmetric configurations. In the following we  use the same conventions as in \cite{gabgre}.\\
In terms of boundary states, the  conservation of dynamical supersymmetries is as in flat space  encoded in
\begin{equation}\label{boundsusy} \left(Q_{\dot a}+i\eta M_{\dot a\dot b}\tilde Q_{\dot b}\right) ||B\rangle\rangle = 0,\end{equation} where for  consistency with  the supersymmetry algebra $M$ has to be an orthogonal matrix.  Assuming  the usual Dirichlet conditions (\ref{bosboundd}) for a flat D-brane which
 in terms of  (closed string) modes reads \begin{equation}\label{dirich} (a^i_n-\tilde a^i _{-n})||B\rangle\rangle=0,\end{equation} the condition (\ref{boundsusy}) uniquely determines the
 fermionic gluing conditions. Using the mode expansion of the dynamical
 supersymmetries given in  appendix  A of \cite{gabgre}, they can be determined  to be of the form \begin{equation}\left( S_n^a+i\eta K_n^{ab} \tilde S^b_{-n} \right)||B\rangle\rangle=0 \;\;\;\;\; (n\neq 0)
 \label{boundferm1}  \end{equation} with \begin{equation}\label{fermglue}  K_n=\frac{1-\frac{\eta m}{2\omega_n c_n^2} \Pi M^t}{ 1+\frac{\eta m}{2\omega_n c_n^2} M\Pi} M=\frac{1-\eta\frac{\omega_n-n}{m}\Pi M^t}{1+\eta\frac{\omega_n-n}{m}M\Pi}M, \end{equation}   where we furthermore used \begin{equation} \left[M, \gamma^i\right]=0\end{equation} for the Dirichlet directions.
The last equation    will be further discussed in the open string
setting later on.
 The formula  (\ref{fermglue}) appeared already in \cite{oblique} in the context of the oblique OD3 brane.\\ In a next step we have  to test  whether the contributions from the Neumann directions in (\ref{boundsusy}) vanish when  appropriate bosonic gluing conditions are enforced.  Using (\ref{boundferm1}) together with the relation  \begin{equation}\label{bosglue} \left(a_n^I-  N^{IJ}_n \tilde a^{J}_{-n}\right) ||B\rangle\rangle=0\end{equation} with a mode depending gluing matrix $N_n$ to be determined below,
 the boundary condition (\ref{boundsusy}) leads to
\begin{eqnarray}0 = \sum_{n\in\mathbb{Z}} \left[  \left( \gamma^I +\frac{m\eta}{2\omega_n c_n^2} M\gamma^I\Pi\right) a^I_{-n} S_n + i\eta \left(M\gamma^J-\frac{m\eta}{2\omega_n c_n^2}\gamma^J\Pi\right) \tilde{a}_n^J\tilde{S}_{-n}   \right] ||B\rangle\rangle\end{eqnarray} which simplifies to
\begin{equation}\left( \gamma^I +\frac{m\eta}{2\omega_n c_n^2} M\gamma^I\Pi\right)N^{IJ}_{-n} K_n- \left(M\gamma^J-\frac{m\eta}{2\omega_n c_n^2}\gamma^J\Pi\right)=0.\end{equation}
Using (\ref{fermglue}) and \begin{equation} \frac{m\eta}{2\omega_n
c_n^2}=\eta \frac{\omega_n-n}{m}\end{equation} this finally gives
\begin{eqnarray} \label{cond1}  &&  M\gamma^J M^t -\eta
\frac{\omega_n-n}{m} \left(\gamma^J\Pi M^t-M \gamma^J
\Pi\right)-\frac{(\omega_n-n)^2}{m^2}\gamma^J \\\nonumber  & = &
N^{IJ}_{-n} \left[ \gamma^I-\eta \frac{\omega_n-n}{m}
\left(\gamma^I\Pi M^t-M\gamma^I
\Pi\right)-\frac{(\omega_n-n)^2}{m^2} M\gamma^I M^t
\right].\end{eqnarray}
Assuming  an $n$-independent gluing matrix $N_n=N$ as for example in \cite{billo}, this would lead to the conditions \begin{equation} M\gamma^J M^t = N^{IJ} \gamma^I;\;\;\;\; \gamma^J= N^{IJ} M\gamma^I M^t\end{equation} and therefore especially to $N^2=1$. As for   consistency of the bosonic gluing conditions $N$ furthermore has to be orthogonal, this gives $N\equiv -1$ for the Neumann directions and thus  we would be  left with  the situation of a vanishing boundary condensate. As in the case of the (4,0) brane studied in \cite{skentay2, gabgre}, a non-vanishing ${\cal F}$ therefore necessarily requires mode depending gluing conditions.\\[2ex] To find the for the present  case  correct gluing matrix $N_n$  we consider  the open string boundary condition (\ref{bosbound}) which gives the open-string operator  identifications \begin{equation} a_n^I= -\left[\frac{{\cal F} +\frac{n}{\omega_n}}{{\cal F} -\frac{n}{\omega_n}}\right]^{IJ} \tilde a_n^J.\end{equation} It is for example  known from the literature on massive integrable boundary field theories  how to translate this directly  into the closed string boundary state  picture \cite{ghoshal}. \\ This so called \emph{crossing} which is usually done as an analytic continuation in the rapidity variable $\theta$ defined as $n=\sinh\theta_n$,  is here simply given by\begin{equation}\theta_n\rightarrow \frac{i\pi}{2}-\theta_n, \;\;\;\;  n\rightarrow i\omega_n,\;\;\;\; \omega_n\rightarrow -i n \end{equation}
which leads to the bosonic gluing conditions \begin{equation}\label{bound3} \left( a_n^I-\left[\frac{{\cal F} -\frac{\omega_n}{n}}{{\cal F} +\frac{\omega_n}{n}}\right]^{IJ}\tilde a_{-n}^J\right) ||B\rangle\rangle=0.\end{equation}
Plugging (\ref{boundN2}) into this and comparing it with  (\ref{bosglue}) gives \begin{equation}\label{bound2}  N_n=\frac{{\cal F}-\frac{\omega_n}{n}}{{\cal F}+ \frac{\omega_n}{n}}=\frac{-(\omega_n-n)+(\omega_n+n)N}{(\omega_n+n)-(\omega_n-n)N}\end{equation} with \begin{equation} N_{-n}=N_n;\;\; N_n^t N_{-n}=N_n^t N_n=1.\end{equation} The last relations  make the bosonic gluing conditions self-consistent.\\ By translating (\ref{bound3}) back into a relation between fields, one  obtains \begin{equation}\label{boundclosed} \left(\partial_\tau x^I-{\cal F}^{IJ} \partial_\sigma x^J  \right)||B\rangle\rangle=0 \;\;\;\; (\tau=0).\end{equation} The   compared to (\ref{bosbound})
additional minus sign  has its origin in the double Wick rotation $\sigma\rightarrow -i\sigma$ and $\tau\rightarrow i\tau$ which effectively takes place when changing from  the open to the closed string channel. \\[2ex] Going with (\ref{bound2}) into (\ref{cond1}) one obtains \begin{eqnarray} 0= (M\gamma^LM^t-\gamma^J N^{JL})+\frac{\eta m}{2n}(\gamma^J\Pi M^t-M\gamma^J\Pi)(N^{JL}-\delta^{LJ})\end{eqnarray} which implies the two following conditions for maximally supersymmetric branes \begin{eqnarray} M\gamma^J M^t=\gamma^I N^{IJ} \\ \left(\delta^{KR}-N^{KR}\right)\left[ \gamma^K\Pi M^t-M\gamma^k\Pi\right]=0.\end{eqnarray} As we are currently only considering  Neumann directions, the  matrix   $1-N$   is invertible, such that  we finally  obtain  the conditions \begin{eqnarray} \label{condN}  M\gamma^J M^t=\gamma^I N^{IJ} \\ \label{condN2} \gamma^K=M\gamma^K\Pi M \Pi= \gamma^I N^{IK} M\Pi M \Pi\end{eqnarray} for  possible gluing matrices of  maximally supersymmetric static branes in the plane wave background.\\[3ex]
So far  we only considered   contributions from  nonzero modes to (\ref{boundsusy}). As the ``zero-modes'' do not contain a $\sigma$-dependency in the closed string channel, ${\cal F}$ drops out for these modes   and  the previous considerations for a vanishing ${\cal F}$ for example in \cite{gabgre}  remain unaltered:  Commuting (\ref{boundsusy}) with $x_0^I$ one   obtains \begin{equation} \left(S_0^a+i\eta M^{ab} \tilde{S}_0^b\right) ||B\rangle\rangle=0,\end{equation} that is, the boundary state preserves 8 kinematical supersymmetries. Using this in (\ref{boundsusy}), we are   left with \begin{equation} \left(-i\eta P_0^I(\gamma^I M-M\gamma^I)\tilde{S}_0-mx_0^I(\gamma^I\Pi-M\gamma^I\Pi M)\tilde{S}_0    \right) ||B\rangle\rangle=0.\end{equation} For the Neumann directions this is solved with the standard requirement \begin{equation} P_0^I ||B\rangle\rangle=0.\end{equation} For the Dirichlet directions, however, one has to have either \begin{equation} M\Pi M\Pi=1,\end{equation} which  corresponds  to a \emph{class II} brane without gauge field excitations or \begin{equation} x_0^i ||B\rangle\rangle=0,\end{equation} that is, the brane has to  be restricted to the origin in transverse space.

\section{Supersymmetric Branes with nontrivial ${\cal F}$} \label{solutions}
The first condition (\ref{condN}) is identical to (\ref{triality1}) for branes in  flat space. As already mentioned above, it says that the 3 matrices $M_{\dot a\dot b}, M_{ab}$ and $N_{IJ}$ are related by $SO(8)$-triality, compare  for example with \cite{gregut}.  This condition is explicitly solved by the formulas given in \cite{gregut, gregut2} (with a slightly different normalization)
\begin{equation} N_{IJ}= e^{\frac{1}{2} \Omega_{MN}\Sigma^{MN}_{IJ}}\end{equation} and \begin{equation}\label{matrixM}  M_{\dot a\dot b}=e^{\frac{1}{4}\Omega_{MN}\gamma^{MN}_{\dot a\dot b}};\;\;\;\; M_{ ab}=e^{\frac{1}{4}\Omega_{MN}\gamma^{MN}_{ab}}\end{equation} with \begin{equation} \Sigma_{IJ}^{MN}=\delta^M_I\delta_J^N-\delta^N_I\delta^M_J;\;\; \gamma^{mn}=\frac{1}{2}\gamma^{[m}\gamma^{n]}.\end{equation}  The second condition  (\ref{condN2})  has no flat space analogue as it explicitly contains the matrix $\Pi$.
 For maximally supersymmetric branes this condition gives rise to some qualitative  differences compared to flat space, where a nonzero boundary condensate does not give  rise  to any new constraints.\\[2ex]  Before studying cases with nonzero magnetic fields  in  detail, it is easy to see that all the  considerations so far  are consistent with the previous works on branes in the plane wave background.  Assuming a mode independent fermionic gluing condition  as  for example in \cite{billo,bgg},  one needs $M\Pi M\Pi=-1$    which furthermore gives with (\ref{condN2})  $N=-1$, such that the bosonic gluing conditions finally  reduce to the usual  $N_n=N=-1$.\\ The maximally supersymmetric  class II branes with $M \Pi M \Pi =1$ are not contained in the previous discussion as the (0,0)-instanton does not have Neumann directions and as the (4,0), (0,4) branes
 couple necessarily to the flux ${\cal F}^{+I}$ (\cite{skentay1, takayanagi}), a possibility to be discussed in the context of the $(4,2)$-brane with boundary condensate later on.\\
  The previous  discussion shows in
  addition that the compared to flat space   new feature of mode dependent gluing conditions as in (\ref{boundferm1}) is actually generic  for
  (maximally supersymmetric) branes in the plane wave  and not a speciality of the D-instanton.

\subsection{The (2,0), (0,2)  branes}
 The   cases  of the (2,0) or (0,2)\footnote{For branes with nonzero magnetic fields we use the same labelling as for their ${\cal F}\rightarrow 0$ limits from \cite{skentay2, gabgre}.} (Euclidean D1-)  branes  are solved as follows.  Without loss of generality we choose  the first two coordinates $I=1,2$ as Neumann directions. With this  we obtain \begin{equation}\label{boundN4}  N= \exp\left[ \theta \left( \begin{array}{cc} 0 & 1 \\ -1 & 0 \end{array}\right)\right]=\left(\begin{array}{cc} \cos\theta & \sin\theta\\ -\sin\theta& \cos\theta\end{array}\right)\end{equation} and \begin{equation} M_{ab}=\exp\left[\frac{\theta}{2} \gamma^{12}_{ab}\right]=1_{ab}\cos\frac{\theta}{2}+\gamma^{12}_{ab} \sin\frac{\theta}{2}. \end{equation}
Furthermore we have \begin{equation} \left[M, \Pi\right]=0,\end{equation}
such that the second equation in (\ref{condN}) reads \begin{equation} \gamma^J=\gamma^I N^{IJ} M^2=M\gamma^J M \leftrightarrow M^t\gamma^J=\gamma^J M\;\;\;\; (J=1,2).\end{equation} For the (2,0) case this is an identity without  further conditions on $N$. Thus we can have   arbitrary constant boundary condensates  ${\cal F}$ on  (Euclidean)  D1-branes without  additional supersymmetry breaking.
\\ As it should,  the boundary condensate ${\cal F}$ somehow interpolates between the usual class I (2,0) brane and the (class II)  (0,0) instanton. Choosing $\theta=0$   the   gluing  matrices reduce to   \begin{equation} K_n= \frac{\omega_n-\eta m \Pi}{n};\;\;\; N_n=1,\end{equation}  which  are the conditions for the D-instanton and for $\theta=\pi$  we obtain \begin{equation} K_n= \gamma^{12};\;\;\; N_n=-1,\end{equation} which are the conditions for the usual (2,0) brane.\\ The boundary state of the (2,0) brane, its consistency  with the open string channel description  and further results from the interpolation to the instanton will be  discussed in the next section.

\subsection{The (3,1), (1,3) branes}
Unlike  the flat space case it is in the plane wave background
impossible to turn on a  boundary condensate on  a (true) subspace
of the brane world volume  and  still  maintain   maximal
supersymmetry. This follows simply from the observation that if
$N$ has an eigenvalue $-1$, the condition (\ref{condN2})
immediately leads to the condition of class I branes and
especially to  $N\equiv -1$.\\ Even for a   non-degenerate ${\cal
F}$ the condition  (\ref{condN2}) is in general  not  solvable as the example of
the (3,1) brane  shows\footnote{The following analysis extends
immediately to the case of (4,2), (2,4) branes without additional
${\cal F}^{+I}$ condensates.}. To study this  case it is
convenient to choose a coordinate systems such that the
antisymmetric $\Omega$ in (\ref{matrixM}) takes a particularly
simple form. From the $SO(4)\times SO(4)$ background symmetry we
have in this case only  a $SO(3)$ symmetry on the world volume
which allows us to bring $\Omega$ to the form  \begin{equation}
\Omega=\left( \begin{array}{cccc}0&a&0&0\\ -a&0&0&b\\ 0&0&0&c\\
0&-b&-c & 0     \end{array}\right).\end{equation}(To
block-diagonalise $\Omega$, that is,  to set $b=0$,  a full
$SO(4)$-rotation would generally be necessary.) By using
(\ref{condN2}) we can now show that a nontrivial boundary
condensate on a flat $(3,1)$-brane is not consistent with maximal
supersymmetry (as long as we do not change the gluing conditions
(\ref{dirich}) to allow  non-static configurations). Indeed, using
$\Omega$ as  above and aligning the brane along the  1,2,3,5
directions, we have \begin{equation}
M=\exp\left[{\frac{a}{2}\gamma^{12}+\frac{b}{2}\gamma^{25}+\frac{c}{2}\gamma^{35}}\right].\end{equation}
Evaluating (\ref{condN2}) in the $I=5$ direction  gives $1=\Pi M^2
\Pi=M^2$ which leads  with (\ref{condN}) to  $ N^2=1$, that is,
the case of a trivial boundary condensate ${\cal F}=0$.\\ This
observation, i.e. that  it is impossible to turn on a gauge field
condensate on a (3,1) brane  and still maintain maximal supersymmetry,
is  related to  the previous observation that the class I  branes
break all dynamical supersymmetries when removed from the origin
of transverse space \cite{billo, dabholkar}. Indeed, from the open string bosonic mode expansion (\ref{neumann}) to be derived below, it can be  seen  that the zero modes along the Neumann directions with nontrivial ${\cal F}$   tend in the (well behaved) ${\cal F}\rightarrow \infty \Leftrightarrow \theta\rightarrow 0$  limit to the Dirichlet zero modes describing a brane removed from the origin. From this it follows that only static (Euclidean)  branes related to the instanton (as discussed before)  or the $(4,0),(0,4)$ branes  (to be considered in the next section)  can preserve 8 dynamical supersymmetries when  boundary condensates  are  switched on.
\subsection{The $(4,2)$, $(2,4)$ branes with flux}
 As a second example of a brane with nontrivial gauge condensate ${\cal F}^{IJ}\neq 0$ we will consider in
 this section the $(4,2)$ case which is connected by the limiting process discussed above to  the class II $(4,0)$
  brane. As explained in \cite{skentay1},  the $(4,0)$ brane couples to the nontrivial $F_5$ background flux in a way that the boundary condensate ${\cal F}^{+I}$ is necessarily switched on to obey the   equations of motion. The ${\cal F}^{+I}$-coupling alters  the bosonic gluing conditions along the $I=1,..4$ Neumann directions, but leaves the other gluing conditions as discussed before  in section \ref{gluing}. \\[1ex]
We will start with  a (4,2) - brane and switch on a boundary
condensate along the $A=5,6$ Neumann directions. Using the
fermionic gluing matrix  \begin{equation}
\Mo=\Pi\exp\left[\frac{\theta}{2}\gamma^{56}\right]\end{equation}
and employing the gluing conditions (\ref{dirich}) and
(\ref{fermglue}) as before,  the  condition (\ref{boundsusy})
  \begin{equation} \left(Q_{\dot a}+i\eta M_{\dot a\dot b}\tilde Q_{\dot b}\right) ||(4,2),{\cal F}^{+I}, \theta \rangle\rangle = 0\end{equation}  uniquely determines the bosonic gluing conditions along the $I=1,\dots, 4$ directions to\footnote{We will give a more detailed derivation of this result in the open string setting in section \ref{open}.}
 \begin{equation} \left(\partial_+ X^I+\partial_- X^I-im\cos\frac{\theta}{2}X^I\right)||(4,2),{\cal F}^{+I}, \theta \rangle\rangle  = 0\;\;\;(\tau=0)\end{equation}
 which in terms of modes reads
\begin{equation}\label{bos235}
\left[\overline{a}^I_0+\frac{1-\cos\frac{\theta}{2}}{1+\cos\frac{\theta}{2}}a_0^I\right]
||(4,2),{\cal F}^{+I}, \theta \rangle\rangle = 0\end{equation} and
\begin{equation}\label{bos234}\left[a^I_n+\left(\frac{\omega_n-m\cos\frac{\theta}{2}}
{\omega_n+m\cos\frac{\theta}{2}}\right) \widetilde{a}^I_{-n}
\right] ||(4,2), {\cal F}^{+I}, \theta\rangle\rangle =
0.\end{equation} The gluing conditions (\ref{bos234}) are  in
direct analogy to the (4,0) case discussed for example in
\cite{skentay2,gabgre}. As for the case of the (2,0) brane,  the
boundary condensate ${\cal F}$ interpolates smoothly between the
(4,2) and the (4,0) brane. It is worth noting that in the
$\theta\rightarrow \pi$ limit not only ${\cal F}^{AB}$,  but also
${\cal F}^{+I}$ tends to zero to exactly reproduce the class I
setting of \cite{billo, skentay2}.

\section{Boundary states and cylinder diagrams} \label{boundarystates}
In this section the (2,0) brane boundary state will be determined. By using it  (and the analogous state for a (4,2)  brane  with ${\cal F}\neq 0$), certain cylinder diagrams will be found  by calculating the corresponding boundary state overlaps, generalizing the results of \cite{bgg, gabgre}. 

\subsection{The (2,0) boundary state}
From the gluing conditions derived in the last section and the standard (anti-) commutation relations summarized for example in the appendix of \cite{gabgre},  one can immediately write down the boundary state of the (2,0) brane with boundary condensate $\cal F$  at  transverse position ${\bf y}={\bf 0}$ (compare for example  with \cite{billo, gabgre, skentay4}) \begin{eqnarray}\nonumber ||(2,0), {\bf 0}, \eta, P^+\rangle\rangle =  {\cal N}^{(2,0)}_{\theta}\exp\left[\sum_{k=1}^\infty \left(\frac{1}{\omega_k} a_{-k}^i\tilde{a}_{-k}^i+ \frac{N^{IJ}_{k}}{\omega_k} a_{-k}^I\tilde{a}_{-k}^J-i\eta K_k^{ab} S^a_{-k} \tilde{S}^{b}_{-k}     \right) \right]\\
\exp\left[ -\frac{1}{2} \left[\frac{1+\eta M}{1-\eta M}\right]_{ab} \theta^a_L\theta^b_L-\frac{1}{2} \left[\frac{1-\eta M}{1+\eta M}\right]_{ab}\overline{ \theta}^a_R\overline{\theta}^b_R  \right]   e^{\frac{1}{2} a_0^i a_0^i-\frac{1}{2} a_0^I a_0^I}|0\rangle. \label{boundarys} \end{eqnarray}
The state $|0\rangle$ is here  given by  the usual Fock space vacuum corresponding to the fixed light-cone momentum $P^+$. It is in particular annihilated by the fermionic zero modes $\overline{\theta}_L$ and $\theta_R$, as defined in the App. A of \cite{gabgre}. The  normalization factor ${\cal N}_{\theta}^{(2,0)}$   has still   to be identified by a comparison with the open string one-loop calculation to be carried out in section 6.
\subsubsection{Cylinder Diagrams} As described for example
in \cite{bgg}, the cylinder diagram is given in terms of  boundary states  by the following
 overlap \begin{equation}{\cal A}_{\overline{\eta}, \eta,\theta}=  \langle\langle(2,0), {\bf 0},
 \overline{\eta}, -P^+, \theta|| e^{-2\pi t H P^+} || (2,0), {\bf 0}, \eta, P^+, \theta\rangle\rangle .\end{equation}
 Keeping in mind the different momenta $P^+$ for the in-  and out-going boundary states and  its consequences,
   the overlap can be evaluated by  standard methods. One obtains for the brane/brane case $\eta=\overline{\eta}$
   \begin{equation}\label{aetaeta} {\cal A}_{\eta,\eta, \theta}=
    \left({\cal N}_{\theta}^{(2,0)}\right)^* { \cal N}_{\theta}^{(2,0)}\end{equation}
     and for the brane/antibrane case $\overline{\eta}=-\eta$
     \begin{equation}\label{aeta-eta} {\cal A}_{\eta, -\eta, \theta}= \frac{\left({\cal N}_{\theta}^{(2,0)}\right)^* { \cal N}_{\theta}^{(2,0)}   }{\left(2\sinh\left[m\pi\sin\frac{\theta}{2}  \right]\right)^4} \frac{(g^{(m)}_2(q, \theta))^4}{(f^{(m)}_1(q))^8}, \end{equation} where $f^{(m)}_1(q)$ is defined as in \cite{bgg} and $g^{(m)}_2(q,\theta)$ is the following deformation of the function $g_2^{(m)}(q)$ of \cite{gabgre}: \begin{eqnarray}\nonumber g_2^{(m)}(q, \theta)&=& 2\sinh\left[m\pi\sin\frac{\theta}{2}  \right] q^{-2\Delta_m}   \sqrt{\left(1+\frac{\sin^2\frac{\theta}{2}}{(1-\cos\frac{\theta}{2})^2}\;  q^m \right)\left(1+\frac{\sin^2\frac{\theta}{2}}{(1+\cos\frac{\theta}{2})^2}\; q^m \right)}\\& & \prod_{n=1}^{\infty} \left(1+q^{\omega_n} \frac{\omega_n-m\cos\frac{\theta}{2}}{\omega_n+m\cos\frac{\theta}{2}} \right) \left(1+q^{\omega_n} \frac{\omega_n+m\cos\frac{\theta}{2}}{\omega_n-m\cos\frac{\theta}{2}} \right)\label{g2theta1}\end{eqnarray}
\begin{equation} \label{g2theta} = 2\sinh\left[m\pi\sin\frac{\theta}{2}  \right]
 q^{-2\Delta_m} \prod_{n\in \mathbb{Z}}\sqrt{ \left(1+q^{\omega_n}
  \frac{\omega_n-m\cos\frac{\theta}{2}}{\omega_n+m\cos\frac{\theta}{2}} \right)
  \left(1+q^{\omega_n} \frac{\omega_n+m\cos\frac{\theta}{2}}{\omega_n-m\cos\frac{\theta}{2}} \right)  }.  \end{equation}
Here the  zero mode contributions of (\ref{g2theta1}) might alternatively be written as
\begin{eqnarray}  q^{-2\Delta_m}\frac{4\sinh\left[m\pi\sin\frac{\theta}{2}\right]}{\sin\frac{\theta}{2}} \sqrt{\left(\sin^2\frac{\theta}{4}+\cos^2\frac{\theta}{4} q^m\right)\left(\cos^2\frac{\theta}{4}+\sin^2\frac{\theta}{4}q^m\right)},  \end{eqnarray} which will be used below to study the $\theta\rightarrow 0$ limit.\\
Using the open-string result (\ref{zetaeta}) to  be derived later on,  the boundary state normalization factor ${\cal N}_{\theta}^{(2,0)}$ is given up to a phase by \begin{equation}{\cal  N}_{\theta}^{(2,0)}=2\sinh\left[m\pi\sin\frac{\theta}{2}\right],\end{equation} which again reproduces  the $(2,0)$ result of \cite{bgg}, but vanishes in the instanton limit. This is, however,
not surprising, as the fermionic part of the boundary state (\ref{boundarys}) diverges in this limit to give
altogether a smooth behavior of the different overlaps in both limiting cases. \\
 The behavior under modular transformations of this $\theta$-dependent family of functions will be discussed
 in the appendix. It can be seen that this family  connects the functions $f_2^{(m)}(q)$ and $g_2^{(m)}(q)$ defined in \cite{bgg, gabgre}: \begin{eqnarray} \lim_{\theta\rightarrow 0}g_2^{(m)}(q,\theta)&=&g_2^{(m)}(q)\\ \lim_{\theta\rightarrow \pi}g_2^{(m)}(q,\theta)&=& 2\sinh\left[m\pi\right]\left(f_2^{(m)}(q)\right)^2\end{eqnarray} and that (\ref{aetaeta}) and (\ref{aeta-eta}) reproduce the (closed string)  results of \cite{bgg,gabgre}.\\[2ex]

\subsection{(4,2)-(0,2) - overlap}

 As an  example of an  overlap containing the (4,2) boundary state
   with nonzero fluxes ${\cal F}^{AB}$ $(A,B =5,6)$ and ${\cal F}^{+I}$ $(I=1\dots 4)$ we consider here
   the cylinder diagram with a (0,2) - (anti-) brane with the same gauge field strength ${\cal F}^{AB}$ on
   its world-volume: \begin{eqnarray} {\cal B}_{\eta,\theta}&=&   \langle\langle(0,2), {\bf 0}, {\eta}, -P^+, \theta||
    e^{-2\pi t H P^+} || (4,2), {\bf 0}, P^+,\theta, {\cal F}^{+I}\rangle\rangle. \end{eqnarray}
     For the overlap with the (0,2) - brane
    ($\eta=1$) we find \begin{eqnarray} {\cal B}_{(\eta=1),
    \theta}&=&
    {\cal N}_\theta^{(0,2)}{\cal N}^{(4,2)}_\theta
  \left(  \frac{\prod_{n=1}^{\infty} \left( 1- q^{\omega_n}\right)}{\prod_{n=1}^{\infty} \left(
  1+\frac{\omega_n-m\cos\frac{\theta}{2}} {\omega_n+m\cos\frac{\theta}{2}}
  q^{\omega_n}\right)}\right)^4
  \end{eqnarray}
and for the antibrane ($\eta=-1$)
\begin{eqnarray} {\cal B}_{(\eta=-1),
    \theta}&=&
    {\cal N}_\theta^{(0,2)}{\cal N}^{(4,2)}_\theta
    \left(\frac{\prod_{n=1}^{\infty} \left(
  1+\frac{\omega_n+m\cos\frac{\theta}{2}} {\omega_n-m\cos\frac{\theta}{2}}
  q^{\omega_n}\right)}{\prod_{n=1}^{\infty} \left( 1-
  q^{\omega_n}\right)}\right)^4,
\end{eqnarray}
where the zero-mode contributions as for example from the bosons
$(I=1,.., 4)$ \begin{equation}
\left(1+\frac{1-\cos\frac{\theta}{2}}{1+\cos\frac{\theta}{2}}q^m\right)^{-2}
\end{equation} cancel  out with  the corresponding fermionic
contributions.\\
Besides the product representation of the $f_1^{(m)}-$function  from
\cite{bgg}, the appearing special functions are essentially given
by  different halves of the function $g_2^{(m)}$ defined in
(\ref{g2theta}), which itself have a good behavior under modular
transformations.  A comparable result was found in \cite{gabgre}
for the (4,0)-(2,0) overlap  without gauge-field excitations.
\section{Open string description} \label{open}
In this section we will study the previously mentioned branes from the open string point of view. First, we will consider  open string (dynamical) supersymmetries  in general and reproduce the conditions (\ref{condN}) and (\ref{condN2}) derived beforehand in the   boundary state approach. After that, we will give a detailed  treatment  of open strings in between   $(2,0)$ branes with ${\cal F}\neq 0$ and compare the results with  those  from the  closed string picture. \\[2ex]
The bosonic and fermionic open string  equations of motion  are the same as in the closed string sector, that is \cite{metsaev1, metsaev2}
\begin{equation} \left(\partial_+\partial_-+\mh^2\right)X^s=\left(\partial_\tau^2-\partial_\sigma^2+\mh^2\right)X^s=0 \label{eombosons} \end{equation} for the bosons  and \begin{equation} \label{eomfermions} \partial_+S=\mh\Pi \widetilde{S};\;\;\; \partial_-\widetilde{S}=-\mh\Pi S\end{equation} for the fermions. As explained in detail in  \cite{bgg, gabgre}, the  open string mass parameter  $\mh$ appropriate for the  light-cone gauge description of instantonic branes is here given by  \begin{equation} \mh=\mu X^+\end{equation} instead of  \begin{equation} m = 2\pi\mu P^+\end{equation} which is  used in the closed string sector. The  bosonic  boundary conditions for the case of a non-vanishing boundary condensate ${\cal F}$ are as described in section \ref{review}  \begin{equation} \partial_\sigma X^I+{\cal F}^{IJ}\partial_\tau X^J=0;\;\;\;\; \sigma=0,\pi\end{equation} for the Neumann  and  \begin{equation} X^i=y^i_{\sigma}\;\;\;\;\ \sigma=0,\pi\end{equation} for the Dirichlet directions. For the fermions we use  furthermore  \begin{equation}\label{boundfz} S(\tau, \sigma=0)=\Mo\widetilde{S}(\tau, \sigma=0),\;\;\;\;\;\; S(\tau, \sigma=\pi)=\eta \Mo\widetilde{S}(\tau, \sigma=\pi)\end{equation} where  $\eta=\pm 1$ distinguishes between the case of a brane -  brane or a brane - antibrane pair.
\subsection{Open string supersymmetries}
The dynamical supersymmetries in the closed string sector follow from the  conserved currents  \cite{metsaev1, skentay2}
\begin{eqnarray}\label{qopen1} Q^\tau &=& \partial_-X^s \gamma^s S-m X^s\gamma^s\Pi \widetilde{S}\\ Q^{\sigma} &=& \partial_- X^s\gamma^s S+mX^s\gamma^s\Pi \widetilde{S}\\\label{qopen2} \widetilde{Q}^\tau &=&\partial_+X^s\gamma^s\widetilde{S}+mX^s\gamma^s\Pi S \\ \widetilde{Q}^{\sigma}&=&-\partial_+ X^s\gamma^s\widetilde{S}+mX^s\gamma^s\Pi S\end{eqnarray} from which the (conserved) supercharges are obtained in the usual way to, for example,   \begin{eqnarray}\label{qclosed1} \sqrt{2P^+}Q_{\dot{\alpha}}&=& \frac{1}{2\pi}\int_0^{2\pi} d\sigma \left(\partial_-{X}^s\gamma^s S-mX^s\gamma^s\Pi \widetilde{S}\right). \end{eqnarray} The conserved supersymmetries in the open string sector are deduced from this as a suitable  linear combination \begin{equation} \label{qopen} Q_{\text{open}} = Q- K \widetilde{Q}\end{equation} of the for open strings  generally time-{\em dependent} charges $Q$ and $\widetilde{Q}$ with a so far undetermined constant $SO(8)$-spinor matrix $K$.\\ From the definition of (\ref{qopen}) in terms of  (\ref{qopen1}) and (\ref{qopen2}) it follows immediately that the open string supercharges (\ref{qopen}) become time independent  iff  the following boundary terms vanish
\begin{equation}\label{boundterm} \left.  \left( \left[\partial_- X^s\gamma^s S+mX^s\gamma^s\Pi \widetilde{S}\right]-K \left[-\partial_+ X^s\gamma^s\widetilde{S}+mX^s\gamma^s\Pi S\right] \right)\right|_{\sigma=0}^\pi.\end{equation}
 Separating this into  contributions from Neumann and Dirichlet directions and furthermore using the gluing conditions (\ref{bound1}) and (\ref{boundfz}),    we obtain  the conditions \begin{equation} 0= \left. \partial_-X^I\left(\gamma^I\Mo-N^{JI}K \gamma^J\right) \widetilde{S}+mX^I\left(\gamma^I\Pi-K\gamma^I\Pi \Mo\right) \widetilde{S}\right|_{\sigma=0}^\pi\end{equation} from the Neumann  and \begin{equation}0=\left. \partial_-X^i\left(\gamma^i \Mo-K\gamma^i\right) \widetilde{S}+mX^i\left(\gamma^i\Pi-K\gamma^i\Pi \Mo\right)\widetilde{S}\right|_{\sigma=0}^\pi\end{equation} from the Dirichlet directions. These conditions require then in particular  \begin{equation} \gamma^I N^{IJ}= K^t\gamma^J \Mo;\;\; \gamma^I=K^t\gamma^I\Pi \Mo^T\Pi;\;\;
\gamma^i \Mo-K\gamma^i=0.\end{equation} Comparing this with (\ref{condN}) and (\ref{condN2}), it  leads to $K=\Mo=M^t$. As long as $\Pi\Mo\Pi\Mo\neq 1$ (for non - class II branes), we furthermore have to choose $y^i_{\sigma}=0$, that is,  to place the branes at the origin of transverse space  to fulfill the conditions imposed by the Dirichlet directions.\\[2ex]
The previous considerations lead for a (2,0) brane with ${\cal F}\neq 0$  to the use of the following  matrix  \begin{equation} \Mo=M^t=\exp\left[-\frac{\theta}{2}\gamma^{12}\right]\end{equation} in  (\ref{boundfz}).  In the conventions of  \cite{dabholkar, skentay1} this actually corresponds to an interpolation to the  anti (2,0) brane, as  $\Mo\rightarrow -\gamma^{12}$ for $\theta\rightarrow \pi$. But this is simply the  usual sign ambiguity in between the open - and closed string picture quantities and we will  still refer to  this as  the (2,0) brane.\\
 The  conserved supercharges corresponding to the choice of  $\widehat{M}$ as given above again  interpolate  between the instanton and  the (2,0) supercharges appearing respectively in \cite{gabgre} and \cite{dabholkar, skentay2} where the combinations  $Q_{(0,0)}=Q-\widetilde{Q}$  were used for the instanton and $Q_{(2,0)}=Q+\gamma^{12}\widetilde{Q}$ for the $(2,0)$ - brane.

\subsubsection{The $(4,2)$ brane with ${\cal F}^{+I}\neq 0$}
For the open string description of the  $(4,2)$ brane  with boundary condensates  we use the following
Ansatz for the  boundary conditions  at $\sigma = 0, \pi$:
\begin{eqnarray} \partial_+ X^i+\partial_-X^i &=&  0;\; X^i=0; \;\; i=7,8\\ \partial_+ X^A+N^{AB}\partial_- X^B &=&0;\;\;
A, B=5,6\\ \partial_+ X^I-\partial_- X^I+\alpha m X^I &=& 0; \;\;
I=1,\dots 4 \\ S &=& \widehat{M} \widetilde{S}.  \end{eqnarray}
Going with this into  equation (\ref{boundterm}),  one obtains for
the Dirichlet  ($i=7,8$) and the Neumann  ($A=5,6$) directions
exactly the same conditions as in (\ref{condN}) and  (\ref{condN2})
when using $K=\widehat{M}$ in (\ref{qopen}) as before. These
conditions  are in particular solved by the  matrix \begin{equation}
\widehat{M}=\Pi \exp\left[- \frac{\theta}{2}\gamma^{56}\right]
\label{m56} \end{equation} which gives in the $\theta\rightarrow
\pi$ limit the correct gluing matrix for the $(4,2)$ brane (for
which then actually \emph{all} ${\cal F}$ components   will be
seen to vanish).\\ For $I= 1,\dots 4$  the requirement for eight
conserved dynamical supersymmetries becomes \begin{equation}\left.
0= \partial_- X^I \left(\gamma^I
M+M\gamma^I\right)\widetilde{S}+mX^I
M\left(M^t\gamma^I\Pi-\gamma^I\Pi M-\alpha
\gamma^I\right)\right|_{\sigma=0}^\pi\end{equation} which vanishes
with (\ref{m56}) when using  $\alpha=2m\cos\frac{\theta}{2}$. The
last equality  determines the strength of the ${\cal F}^{I+}$
components in dependency of the `transverse' field-strength ${\cal
F}^{AB}$.

\subsection{The (2,0) -  Open String description} In the following we will give a detailed open string
treatment of the (2,0) brane with  nontrivial boundary condensate. First we will derive the relevant bosonic
 and fermionic mode expansions and  determine the light-cone gauge  Hamiltonian. After a brief  discussion of
  the canonical quantization,  we will    finally calculate
   some open string partition functions and relate them to closed string boundary state overlaps calculated beforehand  by  modular transformations. It is worth mentioning that  the analogous  case of a  (4,2) brane  with flux discussed already in the closed string sector,  is dealt with by essentially the same calculations, such that we will omit the details here.
\subsubsection{Bosons} The most general  solution of (\ref{eombosons}) is given by \begin{eqnarray} x^s(\sigma, \tau)=A^s \sin(\mh\tau)+\tilde{A}^s \cos(\mh\tau)+B^s\cosh(\mh\sigma)+\tilde{B}^s\sinh(\mh\sigma)\\ \nonumber +i \sum_{n,\omega_n\in \mathbb{C}\backslash \{0\} }\frac{1}{\omega_n}\left(A_n^s e^{-i(\omega_n\tau-n\sigma)}+\tilde{A}_n^s e^{-i(\omega_n\tau+n\sigma)}   \right)\end{eqnarray} with $\omega_n^2=n^2+\mh^2$. Using this,  the  mode expansion for the Dirichlet directions of strings ending on branes at the origin of transverse space  equals \begin{equation} X^i(\tau, \sigma)=-2\sum_{n\in\mathbb{Z}\backslash\{0\}} \frac{e^{-i\omega_n\tau}}{\omega_n} \alpha_n^i \sin(n\sigma);\;\;\;\;\omega_n=\text{sgn}(n)\sqrt{n^2+\widehat{m}^2}.\end{equation} For the Neumann directions,
 on the other hand, one has \begin{eqnarray}\label{neumann}  X^I(\sigma,\tau)&=& e^{-i\mh\sin\frac{\theta}{2}\tau}\exp\left[iJ  \mh\cos\frac{\theta}{2}\sigma   \right]a^I\\ \nonumber  &+& e^{i\mh\sin\frac{\theta}{2}\tau}\exp\left[-i J  \mh\cos\frac{\theta}{2}\sigma   \right]a^{\dagger I}\\ \nonumber  &+&i \sum_{n\in\mathbb{Z}\backslash\{0\}} \frac{e^{-i\omega_n\tau}}{\omega_n}\left[ \alpha_n^I e^{in\sigma}+\tilde{\alpha}_n^I e^{-in\sigma}\right] \end{eqnarray}
with \begin{equation}J= \left(\begin{array}{cc} 0& 1 \\ -1 & 0
\end{array}\right);\;\;\; \tilde{\alpha}_n^I=-\frac{{\cal
F}-\frac{n}{\omega_n}}{{\cal
F}+\frac{n}{\omega_n}}\alpha_n;\;\;\;\;\; {\cal
F}=\left(\begin{array}{cc} 0& f\\-f& 0
\end{array}\right);\;\;\;\;f=\frac{\cos\frac{\theta}{2}}{\sin\frac{\theta}{2}},\label{bosident1}
\end{equation}  where the last equation follows from comparing
(\ref{boundN2}) with (\ref{boundN4}) which gives \begin{equation}
N=\frac{1}{1+f^2}\left(\begin{array}{cc} -(1-f^2) & 2f \\ -2f &
-(1-f^2)    \end{array}  \right).\end{equation} The first two
terms in (\ref{neumann}) which are the substitute for the
otherwise absent zero modes in the case of a boundary condensate,
correspond to modes $n$ for which  the matrix ${\cal F}\pm
\frac{n}{\omega_n}$ appearing in (\ref{bosident1}) is degenerate,
that is, when \begin{equation} 0=\det \left({\cal F}\pm
\frac{n}{\omega_n}\right)=f^2+(\frac{n}{\omega_n})^2\Leftrightarrow
n=\pm i \mh \cos\frac{\theta}{2}.\end{equation} As already
mentioned in \cite{chu} where this  (bosonic)  mode expansion for
${\cal F}\neq 0$  was  written down in different conventions which
more closely resemble the usual  flat space description  as for
example in \cite{chu2}, but which have a  singular behavior  in
the $\theta\rightarrow 0$ limit, these two extra terms fulfill the
condition (\ref{bosbound}) actually for all $\sigma$, not only on
the boundary.  In the limits $\theta\rightarrow \pi, 0$ these
terms tend to (redefinitions of)  the usual ``zero''-modes of the
$(2,0)$-brane  or the instanton.

\subsubsection{Fermions}
The most general solutions  of the fermionic equations of motion  (\ref{eomfermions}) are  given by \cite{gabgre}

\begin{eqnarray} S(\sigma, \tau)&=& S \cos(\mh\tau)+\Pi \tilde{S} \sin(\mh\tau)+T\cosh(\mh\sigma)+\Pi \tilde{T}\sinh(\mh\sigma)\\ \nonumber &\;\; +& \sum_{n,\omega_m\in \mathbb{C}\backslash\{0\}} c_n\left[S_n e^{-i(\omega_n\tau-n\sigma)}+\frac{i}{\mh}(\omega_n-n) \Pi \tilde{S}_n e^{-i(\omega_n\tau+n\sigma)}\right]\\ \tilde{S}(\tau,\sigma) &=&-\Pi S\sin(\mh\tau)+\tilde{S}\cos(\mh\tau)+T\cosh(\mh\sigma)+\Pi T \sinh(\mh\sigma)\\ \nonumber &\;\; +& \sum_{n,\omega_m\in \mathbb{C}\backslash\{0\}} c_n\left[\tilde{S}_n e^{-i(\omega_n\tau+n\sigma)}-\frac{i}{\mh}(\omega_n-n)\Pi S_n e^{-i(\omega_n\tau-n\sigma)} \right] \end{eqnarray}
with the boundary conditions (\ref{boundfz}) with  $\Mo=\exp\left[-\frac{\theta}{2}\gamma^{12} \right]$.\\[1ex]
 For a nontrivial $\Mo$ with $\theta\in (0,\pi)$ the boundary conditions (\ref{boundfz})  lead in both cases $\eta=\pm 1$ to vanishing `zero'-modes $S=T=\tilde{S}=\tilde{T}=0$. The conditions for the nonzero modes read \begin{eqnarray} \label{condf1} \left( \bbbone +\frac{i}{\mh}(\omega_n-n)\Mo\Pi\right)S_n&=&\left(\Mo-\frac{i}{\mh}(\omega_n-n)\Pi\right) \tilde{S}_n\\  \label{condf2} \left( \bbbone +\eta\frac{i}{\mh}(\omega_n-n)\Mo\Pi\right)S_n&=&\left(\eta \Mo-\frac{i}{\mh}(\omega_n-n)\Pi\right) e^{-2\pi i n} \tilde{S}_n.\end{eqnarray}

\paragraph{brane - brane, $\eta=1$}
Using (\ref{condf1}) and  (\ref{condf2}),  the mode expansions for strings stretching between a brane  - brane pair  can be determined to be
\begin{eqnarray}\label{modesS}  S(\tau,\sigma) &=& \frac{\bbbone+\Pi}{2} \exp\left[ +\mh\sin\frac{\theta}{2} \tau \gamma^{12}\right] S_0 e^{\mh\cos\frac{\theta}{2}\sigma}\\&\;\; +&\frac{\bbbone-\Pi}{2} \exp\left[ -\mh\sin\frac{\theta}{2} \tau\gamma^{12}\right]\Mo S_0 e^{-\mh\cos\frac{\theta}{2}\sigma}\\ &\;\; +&
\sum_{n\in \mathbb{Z}\backslash\{0\}} c_n\left[S_n e^{-i(\omega_n\tau-n\sigma)}+\frac{i}{\mh}(\omega_n-n) \Pi \tilde{S}_n e^{-i(\omega_n\tau+n\sigma)}\right];\end{eqnarray} \begin{eqnarray} \label{modesSt}
\tilde{S}(\tau,\sigma)&=& \frac{\bbbone+\Pi}{2} \exp\left[ +\mh\sin\frac{\theta}{2} \tau \gamma^{12}\right]\Mo^t S_0 e^{\mh\cos\frac{\theta}{2}\sigma}\\&\;\; +&\frac{\bbbone-\Pi}{2} \exp\left[- \mh\sin\frac{\theta}{2} \tau  \gamma^{12}\right] S_0 e^{-\mh\cos\frac{\theta}{2}\sigma} \\
&\;\; +& \sum_{n\in \mathbb{Z}\backslash\{0\}} c_n\left[\tilde{S}_n e^{-i(\omega_n\tau+n\sigma)}-\frac{i}{\mh}(\omega_n-n)\Pi S_n e^{-i(\omega_n\tau-n\sigma)} \right]  \end{eqnarray} with $\omega_n=\text{sgn}(n)\sqrt{n^2+\mh^2}, c_n=\frac{\mh}{\sqrt{2\omega_n(\omega_n-n)}}$, $- 2 \sin\frac{\theta}{2}\gamma^{12}=\Mo-\Mo^t$  and the operator identifications \begin{equation}\label{fermident1}\tilde{S}_n=\frac{\left( \bbbone +\frac{i}{\mh}(\omega_n-n)\Mo\Pi\right)}{\left(\Mo-\frac{i}{\mh}(\omega_n-n)\Pi\right)} S_n.\end{equation} The first two terms in each expansion correspond as in the bosonic case  to $n= \pm i  \mh\cos\frac{\theta}{2}$ for which the matrices in (\ref{condf1}) and (\ref{condf2}) are degenerate. As before,  these ``zero''-modes  actually fulfill the fermionic boundary conditions for all $\sigma\in [0,\pi]$ and not only on the boundary.

\paragraph{brane - antibrane, $\eta=-1$}
Contrary to the situation before,  there are no extra non-zero
mode contribution in  the case of a open string joining a brane
and an antibrane (with the same gauge condensate ${\cal F}$). This
follows directly by combining (\ref{condf1}) and (\ref{condf2}) to
\begin{equation}
\left(n-i\mh\cos\frac{\theta}{2}\Pi\right)\tilde{S}_n=-\left(n+i\mh\cos\frac{\theta}{2}\Pi\right)
e^{-2\pi in}\tilde{S}_n\end{equation} which gives
$\tilde{S}_{n=\pm i\mh \cos\frac{\theta}{2}}=0$.\\ The
identification between the nonzero - modes $\tilde{S}_n$ and $S_n$
is still given by (\ref{fermident1}) which again follows from
(\ref{condf1}). To simultaneously fulfill the second condition
(\ref{condf2}), the moding $n$ has to fulfill the following
equation \begin{equation}\label{modes+} n\in P_\theta^+:\;\;\;\;
\frac{n+i\mh\cos\frac{\theta}{2}}{n-i\mh\cos\frac{\theta}{2}}=-e^{2\pi
in };\;\;\; n\neq 0\end{equation} for the $S^+_n=\frac{\bbbone
+\Pi}{2} S_n$ modes and
\begin{equation}\label{modes-}n\in P^-_\theta:\;\;\;\; \frac{n-i\mh\cos\frac{\theta}{2}}{n+i\mh\cos\frac{\theta}{2}}=-e^{2\pi in };\;\;\; n\neq 0\end{equation} for the $S^-_n=\frac{\bbbone -\Pi}{2} S_n$ modes. This is in direct analogy to the  $(0,0)-\overline{(0,0)}$ situation described in \cite{gabgre}. Both equations have infinitely many solutions on the real axis and for small $\mh$ all of them  are being close to the flat space case of half integers. From $\mh\cos\frac{
\theta}{2}> \frac{1}{\pi}$ on, however, two solutions of $P_\theta^-$ 
become  imaginary. This is somewhat  in analogy to  the
additional `zero-modes' for a string stretching between two
branes of the same kind with nonzero flux as  discussed before.

\subsubsection{The light-cone gauge Hamiltonian}
 The light-cone gauge  Hamiltonian is  given  by \cite{metsaev1, metsaev2} \begin{equation} \frac{X^+}{2\pi}H^{\text{open}}=\frac{1}{4}\int_0^\pi d\sigma\; \left(\dot{X}^2+{X'}^2+\mh^2 X^2\right)+\frac{i}{2}\int_0^\pi d\sigma\; \left(S\dot{S}+\tilde{S}\dot{\tilde{S}}\right)\end{equation} and has the following
 expression in terms of modes
\begin{eqnarray} \label{hamil} \frac{X^+}{2\pi}H^{\text{open}}&=& \frac{\mh }{2\cos\frac{\theta}{2}}\sinh\left[\mh\pi\cos\frac{\theta}{2}\right] \left(a e^{-i\pi\mh J \cos\frac{\theta}{2}} a^\dagger+a^\dagger e^{i\pi\mh J\cos\frac{\theta}{2}} a\right)\\ \nonumber  &+&i \frac{\sin\frac{\theta}{2}}{\cos\frac{\theta}{2}}\sinh\left[\mh\pi\cos\frac{\theta}{2}\right] \left( S_0\frac{1-\Pi}{2}\gamma^{12}S_0 e^{-\mh\cos\frac{\theta}{2}\pi}- S_0\frac{1+\Pi}{2}\gamma^{12}S_0 e^{\mh\cos\frac{\theta}{2}\pi}\right)\\ \nonumber  &+& \pi\sum_{n\neq 0} \left(\alpha_{-n}^I\alpha_n^I+\alpha_{-n}^i\alpha_n^i  +\omega_n S_{-n}S_n\right)\end{eqnarray} for a open string stretching   between a brane - brane pair at the same transverse position ${\bf y}={\bf 0}$. In the brane - antibrane case the fermionic zero modes $S_0$  are absent and the fermionic nonzero modes have to fulfill either (\ref{modes+}) or (\ref{modes-}). In this case there is furthermore a nontrivial normal ordering constant to be discussed briefly  below  which is (apart from possible zero mode contributions) absent in the first case due to supersymmetry.

\subsubsection{Quantization}
The quantization proceeds in the usual way. By  stressing that in our
case ${\cal F}=F, B=0$, it is clear  that in particular the fermionic
canonical conjugated momenta are unaffected by the boundary condensates.
 Requiring therefore the  equal time (anti-) commutation relations
 (as discussed with further details  for the bosons  in \cite{chu, chu2})
\begin{eqnarray} [X^I(\tau, \sigma), P^J(\tau,
\sigma')]&=&i\delta^{IJ} \delta(\sigma-\sigma'),\\  \{S^a(\tau,
\sigma),S^b(\tau, \sigma')\}=\{\tilde{S}^a(\tau,
\sigma),\tilde{S}^b(\tau, \sigma')\}&=&2\pi\delta(\sigma-\sigma')\delta^{ab},\\
\{S^a(\tau, \sigma),\tilde{S}^b(\tau,\sigma')\}& =&0,\end{eqnarray}
one obtains the following relations for the modes:\\ For the
bosons
\begin{eqnarray}\label{bosnonzero}
[\alpha^i_n,\alpha^j_m]&=&\omega_n\delta_{m+n}\delta^{ij}\\{}
[\alpha^I_n,\alpha^J_m]&=&\omega_n\delta_{m+n}\delta^{IJ}
\\{}\label{boszero} [ a^I,a^{\dagger J}]&=&\pi \sin\theta
\left[\frac{\cosh(\mh\pi\cos\frac{\theta}{2})}{\sinh(\mh\pi\cos\frac{\theta}{2})}-i
J\right]^{IJ}\\ &=& \frac{\pi\sin\theta}{\sinh\left[\mh \pi
\cos\frac{\theta}{2}\right]}\exp\left[
-iJ\mh\pi\cos\frac{\theta}{2}\right]^{IJ}\end{eqnarray} and for
the  fermions   \begin{eqnarray} \{S_0^a ,
S_0^b\}&=&\frac{\mh\pi\cos\frac{\theta}{2}}{\sinh\left[\mh\pi
\cos\frac{\theta}{2}\right]}  \left(\frac{1+\Pi}{2} e^{-\pi
\mh\cos\frac{\theta}{2}}+ \frac{1-\Pi}{2} e^{\pi
\mh\cos\frac{\theta}{2}}  \right)^{ab}\label{fermzero} \\
\label{fermnonzero}
\{S_n^a,S_m^b\}&=&\delta_{n+m}\delta^{ab}.\end{eqnarray} Some
details of the derivations will be given in the appendix.
\subsubsection{Partition Functions} In this section the open string partition functions for strings stretching between (2,0) branes with flux will be calculated for the cases  of a brane - brane and a brane - antibrane pair. \\ As discussed in \cite{polchinski, bgg}, these partition functions are given by
\begin{equation}\label{ztt}  Z(\tti)=\text{Tr}\exp\left[-\frac{X^+}{2\pi}H^{\text{open}}\tti\right], \end{equation}
 where the trace runs over the open string Hilbert spaces as (implicitly) determined in the previous subsection.\\[2ex] For the brane - brane pair, the normal ordered contributions of the bosonic zero modes in (\ref{hamil}) are for example  given by
 \begin{equation} \frac{\mh\sinh\left[\mh \pi \cos\frac{\theta}{2}\right]}{\cos\frac{\theta}{2}}
  a^\dagger\exp\left[i\pi\mh J \cos\frac{\theta}{2}\right] a+2\pi\mh\sin\frac{\theta}{2}\end{equation}
   and are therefore  leading  to a factor  \begin{equation} \frac{1}{\left(  2\sinh\left[\mh\pi\tti
    \sin\frac{\theta}{2}\right] \right)^2} \end{equation} as contribution to (\ref{ztt}).\\ Defining
    the fermionic vacuum by the requirement that  it is for example  annihilated by the
    combinations
    $A^l=S_0^{1+2l}+iS_0^{2+2l}$, $l=0,.., 3$,  the fermionic zero modes furthermore give rise to the
    factor \begin{equation} \left(  2\sinh\left[\mh\pi\tti \sin\frac{\theta}{2}\right] \right)^4.\end{equation}
     Together with the nonzero - mode contributions which are evaluated as for example in \cite{gabgre},
       one obtains the following open string partition functions
       \begin{equation}\label{zetaeta} Z_{\eta, \eta, \theta}(\widetilde{t})=
       \left(  2\sinh\left[\mh\pi\tti \sin\frac{\theta}{2}\right] \right)^2\end{equation} and
       \begin{equation}\label{zeta-eta} Z_{\eta, -\eta, \theta}(\widetilde{t}) = \frac{1}{\left(  2\sinh\left[\mh\pi\tti \sin\frac{\theta}{2}\right] \right)^2}\frac{\left(\widehat{g_4}^{(\mh)}(\tti, \theta)\right)^4}{\left(f_1^{(\mh)}(\tti)\right)^8}\end{equation} for open strings stretching between a brane - brane or a brane - antibrane pair. The function $f_1^{(\mh)}(\tti)$ is here defined as in \cite{bgg,gabgre} and we have furthermore   set \begin{equation}\label{gtilde} \widehat{g}^{(\mh)}_4(\tti, \theta)=2\sinh\left[\mh \pi \widetilde{t}\right]\qti^{-\overline{\Delta}_{\mh,\theta}+\frac{\mh}{2}(1-\sin\frac{\theta}{2})}\prod_{\lambda\in P^+_\theta}\sqrt{\left(1-\qti^{\sqrt{\lambda^2+\mh^2}}\right)}\prod_{\lambda\in P^-_\theta}\sqrt{\left(1-\qti^{\sqrt{\lambda^2+\mh^2}}\right)}\end{equation}as $\theta$-dependent generalization of the  function $g_2^{(\mh)}$ appearing in \cite{gabgre}.  The offset $\overline{\Delta}_{\mh, \theta}$ is  essentially determined  by  the  normal ordering constant in the light-cone gauge  Hamiltonian. Its  explicit form will be given in the appendix.\\  As in the closed string picture, this family of functions reproduces the results of \cite{gabgre} in the limits $\theta\rightarrow 0, \pi$: \begin{eqnarray} \lim_{\theta\rightarrow 0}\widehat{g}^{(\mh)}_4(\tti, \theta)&=& \widehat{g}^{(\mh)}_4(\tti)\\  \lim_{\theta\rightarrow \pi}\widehat{g}^{(\mh)}_4(\tti, \theta)&=&2\sinh\left[m\pi\right]\left( f_4^{(\mh)}(\widetilde{q})\right)^2,\end{eqnarray} which is in particular  consistent with the modular transformation properties discussed in \cite{bgg, gabgre}.
\section{Conclusion}  In this paper  we have derived  the general conditions for maximally supersymmetric branes with nontrivial  ${\cal F}^{IJ}$ world-volume fluxes  in the plane wave background.  Both,  the open -  and closed string picture gave rise to the same results, generalizing the findings  of \cite{billo, dabholkar} and \cite{skentay1}.\\ In a next step we have solved these conditions and found  that the constant magnetic boundary  fields give rise to  new (continuous)  families of maximally supersymmetric branes which are connected in the limit of infinite field strengths to the previously classified class II branes.\\ In contradistinction to flat space, magnetic fields cannot be turned on on every maximally supersymmetric brane without breaking some further  supersymmetries. We argued  that this behaviour is directly related to the previous observation from \cite{billo, dabholkar} that class I branes are only supersymmetric when placed at the origin of transverse space.\\  After constructing  boundary states and determining certain  overlaps, we have shown in addition that the new branes pass the important open/closed duality  consistency check by determining open string partition functions and demonstrating their equivalence with the closed string results. This final step involved a new family of modular functions, generalizing  and connecting the results of \cite{bgg, gabgre}.

\section*{Acknowledgements}
It is a pleasure to thank Matthias Gaberdiel
for encouragement and support and Carlo Albert, Stefan
Fredenhagen, Peter Kaste and Martin Weidner for discussions. This
research is partially  supported by the Swiss National Science
Foundation and the European Network `ForcesUniverse' (MRTN-CT-2004-005104).

\begin{appendix}
\section{Appendix}
\subsection{Quantization}
In this section  some details of the canonical quantization for the fermionic and bosonic (open string) degrees of freedom  will be given.
\subsubsection{Fermions}
Using the relation $ \{S_n^a,S_m^b\}=\delta_{n+m}\delta^{ab}$ we obtain with (\ref{modesS}) \begin{eqnarray} \nonumber \{S(\tau, \sigma),S(\tau,\sigma')\} &=& e^{\mh\cos\frac{\theta}{2}(\sigma+\sigma')} \frac{1+\Pi}{2} \{S_0, S_0\}+  e^{-\mh\cos\frac{\theta}{2}(\sigma+\sigma')} \frac{1-\Pi}{2} \{S_0, S_0\}\\ \label{zwischen12}  &+ &\sum_{n\neq 0} \left( e^{in(\sigma-\sigma')} -\frac{\mh i}{2\omega_n}\Pi (K^t_{-n}+K_{-n}) e^{in(\sigma+\sigma')}   \right), \end{eqnarray} where  $K_n$ is the matrix  appearing  in (\ref{fermident1}).  Using \begin{equation} K^t_{-n}+K_{-n}=\frac{2\omega_n\cos\frac{\theta}{2}}{n+i\mh\Pi\cos\frac{\theta}{2}}=2\omega_n\cos\frac{\theta}{2}\;\;\;\frac{n-i\mh\Pi\cos\frac{\theta}{2}}{n^2+\mh^2\cos^2\frac{\theta}{2}}\end{equation} the sum in (\ref{zwischen12}) becomes \begin{equation} \sum_{n\in\mathbb{Z}} \left(  e^{in(\sigma-\sigma')} -i\mh\cos\frac{\theta}{2}\Pi\frac{n e^{in(\sigma+\sigma')} }{n^2+\mh^2\cos^2\frac{\theta}{2}} - \mh^2\cos^2\frac{\theta}{2}\frac{ e^{in(\sigma+\sigma')}}{n^2+\mh^2\cos^2\frac{\theta}{2}}\right).\end{equation}The infinite sums can be evaluated as contour integrals as in \cite{gabgre}, giving for example
\begin{equation}\sum_{n\in\mathbb{Z}}\frac{n e^{in(\sigma+\sigma')} }{n^2+\mh^2\cos^2\frac{\theta}{2}}= -\int_C \frac{1}{1-e^{2\pi i z}}\frac{z}{z^2+\mh^2\cos^2\frac{\theta}{2}} e^{iz(\sigma+\sigma')}\end{equation} where the contour $C$ runs infinitesimally above and below the real axis.  With $0<\sigma+\sigma'<2\pi$ we can close these paths in the upper  / lower half plane, giving
 \begin{equation}
\sum_{n\in\mathbb{Z}}\frac{n e^{in(\sigma+\sigma')} }{n^2+\mh^2\cos^2\frac{\theta}{2}}=\pi i \left(\frac{e^{-\mh\cos\frac{\theta}{2}(\sigma+\sigma')}}{1-e^{-2\pi \mh\cos\frac{\theta}{2}}}+\frac{e^{\mh\cos\frac{\theta}{2}(\sigma+\sigma')}}{1-e^{2\pi \mh\cos\frac{\theta}{2}}}   \right)\end{equation}
and analogously  \begin{equation}
\sum_{n\in\mathbb{Z}}\frac{e^{in(\sigma+\sigma')} }{n^2+\mh^2\cos^2\frac{\theta}{2}}=\frac{\pi}{\mh\cos\frac{\theta}{2}}\left(\frac{e^{-\mh\cos\frac{\theta}{2}(\sigma+\sigma')}}{1-e^{-2\pi \mh\cos\frac{\theta}{2}}}-\frac{e^{\mh\cos\frac{\theta}{2}(\sigma+\sigma')}}{1-e^{2\pi \mh\cos\frac{\theta}{2}}}   \right).\end{equation} Plugging this into (\ref{zwischen12}),  it can be seen with (\ref{fermzero})  that these terms are exactly cancelled by the zero-mode contributions. Altogether this leads to \begin{equation} \{S(\tau, \sigma),S(\tau,\sigma')\}=\sum_{n\in\mathbb{Z}}e^{in(\sigma-\sigma')}=2\pi\delta(\sigma-\sigma');\;\;\; 0<\sigma, \sigma'<\pi,\end{equation} which had to be shown. The remaining  cases follow from a similar analysis.

\subsubsection{Bosons} For the Neumann directions the canonical conjugated  momenta are given by \begin{eqnarray}   P^I&=&\frac{1}{2}\left(\partial_\tau X^I+F^{IJ}\partial_\sigma X^J\right)\\&=&\nonumber
\frac{-i\mh}{2\sin\frac{\theta}{2}} \left(e^{-i\mh\sin\frac{\theta}{2}\tau}\exp[iJ\mh\cos\frac{\theta}{2}\sigma]^{IJ}a^J-e^{i\mh\sin\frac{\theta}{2}\tau}\exp[-iJ\mh\cos\frac{\theta}{2}\sigma]^{IJ}a^{\dagger J}\right)\\\nonumber  &+&\frac{1}{2} \sum_{n\in \mathbb{Z}\backslash \{0\}} e^{-i\omega_n\tau}\left( \left[1-\frac{n}{\omega_n}F\right]^{IJ}\alpha^J_n e^{in\sigma}+\left[1+\frac{n}{\omega_n}F\right]^{IJ}\tilde{\alpha}^J_n e^{-in\sigma}   \right).\end{eqnarray} Using this, (\ref{bosident1}) and $[\alpha^I_n,\alpha^J_m]=\omega_n\delta_{n+m}\delta^{IJ}$,  the nonzero mode contributions to\\ $[X^I(\tau, \sigma),P^J(\tau, \sigma')]$ are given by \begin{equation} i\sum_{n\in \mathbb{Z}} e^{in(\sigma-\sigma')}-i\sum_{n\in \mathbb{Z}}\frac{\mh^2\cos^2\frac{\theta}{2}-n^2}{n^2+\mh^2\cos^2\frac{\theta}{2}} e^{in(\sigma+\sigma')}.\end{equation}  Using the contour integration as before, this leads to \begin{equation} i\delta(\sigma-\sigma')-2\pi i \mh\cos\frac{\theta}{2} \left[\frac{e^{-\mh\cos\frac{\theta}{2}(\sigma+\sigma')}}{1-e^{-2\pi \mh\cos\frac{\theta}{2}}}-\frac{e^{\mh\cos\frac{\theta}{2}(\sigma+\sigma')}}{1-e^{2\pi \mh\cos\frac{\theta}{2}}}    \right].\label{z123} \end{equation} Setting $[a^I,a^{J\dagger}]=L^{IJ}$, the  zero mode contribution are here
\begin{equation}\frac{i\mh}{4\sin\frac{\theta}{2}} \left( e^{\mh\cos\frac{\theta}{2}(\sigma+\sigma')}((1+iJ)L+(1-iJ)L^t)+e^{-\mh\cos\frac{\theta}{2}(\sigma+\sigma')}((1-iJ)L+(1+iJ)L^t)\right).\nonumber\end{equation}  Comparing this with (\ref{z123}) again  uniquely determines $L$ as given in (\ref{boszero}).

\subsection{The modular transformation}
Using the result \begin{equation}
f_2^{(m)}(q)=f_4^{(\mh)}(\qti)\end{equation} from \cite{bgg},
 we have to establish the relation \begin{equation}\label{modular2}
g_2^{(m)}(t,\theta) =
\widehat{g}_4^{(\mh)}(\widetilde{t},\theta)\end{equation} with $\widetilde{t}
= \frac{1}{t}$ for the
special functions defined in (\ref{g2theta}) and (\ref{gtilde}).\\
Setting $m_1=m \cos\frac{\theta}{2}$ the proof given in App. D of
\cite{gabgre} carries over immediately to the present situation,
so that we will not reproduce it here in detail. In that proof the
open string offset $\overline{\Delta}_{\mh, \theta}$ is determined
to \begin{equation} \label{offset} \overline{\Delta}_{\mh,
\theta}=-\frac{1}{(2\pi)^2}\sum_{p=1}^{\infty}
(-1)^p\sum_{r=0}^\infty c_r^p\; \mh \left(\frac{\partial}{\partial
\mh ^2 }  \right)^r  \frac{1}{\mh} \int_0^\infty ds
\left(\frac{-s}{\pi^2\cos^2\frac{\theta}{2}}\right)^r
e^{-p^2s-\frac{\pi^2 \mh^2}{s}}, \end{equation} where the
coefficients $c_r^p$ are taken from the power series expansion
\begin{equation} \label{series1}
\left(\frac{\omega_n+m_1}{\omega_n-m_1}  \right)^p + \left(
\frac{\omega_n-m_1}{\omega_n+m_1}  \right)^p=\sum_{r=0}^\infty
c_r^p \left(\frac{\omega_n}{m_1}\right)^r.\end{equation} As it
should, the limit $\theta\rightarrow 0$ reproduces the instanton
result. The limit $\theta\rightarrow \pi$, however,  is singular, 
as the expansion (\ref{series1}) as used in the derivation of
(\ref{offset})  strictly makes sense  only when
understood  as an analytic continuation for the meromorphic
function on the left hand side  of (\ref{series1}) to values
\begin{equation} m_1=m \cos\frac{\theta}{2}<\omega_n = \sqrt{n^2+m^2}.\end{equation} Doing this,  only the $r=0$
term in (\ref{offset}) contributes with a factor $c_0^p\rightarrow
2$, leading to \begin{equation} \overline{\Delta}_{\mh,
\theta}\rightarrow 2 \Delta'_m\end{equation} with\begin{equation}
\Delta'_m=-\frac{1}{(2\pi)^2}\sum_{p=1}^\infty (-1)^p
\int_0^\infty ds\; e^{-p^2 s-\frac{\pi^2m^2}{s}}\end{equation} in
accordance with the results of \cite{bgg}.\\ As in the last step,
parts of the proof  of (\ref{modular2}) make use of an analytic
continuation in $m_1$. It would be interesting to see whether
there is a more direct approach as for example found by Gannon in
\cite{gannon} for the $f_i^{(m)}$ functions of \cite{bgg}, where  
certain $\theta$-function identities are being used.

\end{appendix}

\end{document}